\documentclass[11pt]{article}
\usepackage{amssymb,latexsym,amsmath}
\usepackage{amsfonts, graphics}
\newtheorem{theorem}{Theorem}
\newtheorem{lemma}{Lemma}
\newtheorem{proposition}{Proposition}

\begin{document}
\title{On static shells and the Buchdahl inequality for the spherically symmetric Einstein-Vlasov system}
\author{H{\aa}kan Andr\'{e}asson\\
        Department of Mathematics, Chalmers,\\
        S-41296 G\"oteborg, Sweden\\
        email\textup{: \texttt{hand@math.chalmers.se}}}

\maketitle

\begin{abstract}
In a previous work \cite{An1} matter models such that the energy density $\rho\geq 0,$ and 
the radial- and tangential pressures $p\geq 0$ and $q,$ satisfy 
$p+q\leq\Omega\rho, \;\Omega\geq 1,$ were considered in the context of 
Buchdahl's inequality. It was proved that static shell solutions of 
the spherically symmetric Einstein equations obey a Buchdahl type inequality whenever 
the support of the shell, $[R_0,R_1], \;R_0>0,$ satisfies $R_1/R_0<1/4.$ Moreover, given a 
sequence of solutions such that $R_1/R_0\to 1,$ then the limit supremum of $2M/R_1$ 
was shown to be bounded by $((2\Omega+1)^2-1)/(2\Omega+1)^2.$ 
In this paper we show that the hypothesis that $R_1/R_0\to 1,$ can be realized 
for Vlasov matter, by constructing a sequence of static shells of 
the spherically symmetric Einstein-Vlasov system with this property. 
We also prove that for this sequence not only the limit supremum of $2M/R_1$ is bounded, 
but that the limit is $((2\Omega+1)^2-1)/(2\Omega+1)^2=8/9,$ since $\Omega=1$ 
for Vlasov matter. Thus, static shells of Vlasov matter can have $2M/R_1$ arbitrary 
close to $8/9,$ which is interesting in view of \cite{AR2}, where numerical evidence 
is presented that $8/9$ is an upper bound of $2M/R_1$ of any static solution of 
the spherically symmetric Einstein-Vlasov system. 
\end{abstract}
\section{Introduction}
Under the assumption of isotropic pressure and non-increasing energy density outwards, 
Buchdahl \cite{Bu1} has proved that a spherically symmetric fluid ball satisfies 
\[
\frac{2M}{R_1}\leq \frac{8}{9},
\]
where $M$ and $R_1$ is the total ADM mass and the outer boundary of the fluid ball respectively. In \cite{An1} Buchdahl's inequality was investigated for spherically symmetric static shells with support in $[R_0,R_1], \;R_0>0,$ for which neither of Buchdahl's hypotheses hold. We refer to the introduction in \cite{An1} for a review on previous results on Buchdahl type inequalities. The matter models considered in \cite{An1} were assumed to have non-negative energy density $\rho$ and pressure $p,$ and to satisfy the following inequality 
\begin{equation}
p+q\leq\Omega\rho,\;\Omega\geq 1,\label{matterhyp} 
\end{equation}
where $q$ is the tangential pressure. 
It was shown that given $\epsilon<1/4,$ there is a $\kappa>0$ such that any static solution of the spherically symmetric Einstein equations satisfies 
\[
\frac{2M}{R_1}\leq 1-\kappa. 
\]
Furthermore, given a sequence of static solutions, indexed by $j,$ with support 
in $[R_0^j,R_1^j]$ where 
\[
R_1^j/R_0^j\to 1 \mbox{ as } j\to\infty,
\]
it was proved that 
\begin{equation}
\limsup_{j\to\infty}\frac{2M^j}{R_1^j}\leq\frac{(2\Omega+1)^2-1}{(2\Omega+1)^2},\label{bound}
\end{equation}
where $M^j$ is the corresponding ADM mass of the solution with index $j.$ 
The latter result is motivated by numerical simulations \cite{AR2} of the spherically symmetric Einstein-Vlasov system. For Vlasov matter $\Omega=1$ and the inequality (\ref{matterhyp}) is strict, and the bound in (\ref{bound}) becomes $8/9$ as in Buchdahl's original work. We will see that for Vlasov matter, a sequence can be constructed such that $R_1^j/R_0^j\to 1,$ and such that the value $8/9$ of $2M/R_1$ is attained in the limit. It should be emphasized that the static solution which attains the value $8/9$ in Buchdahl's case is an isotropic solution with constant energy density, whereas the limit state of the sequence that we construct is an infinitely thin shell which has $p^j/q^j\to 0$ as $j\to\infty,$ which means that it is highly non-isotropic. 

Before describing in more detail the numerical results in \cite{AR2}, 
which provides the main motivation for this work, let us first introduce 
the spherically symmetric Einstein-Vlasov system. 

The metric of a static spherically symmetric spacetime takes 
the following form in Schwarzschild coordinates 
\begin{displaymath}
ds^{2}=-e^{2\mu(r)}dt^{2}+e^{2\lambda(r)}dr^{2}
+r^{2}(d\theta^{2}+\sin^{2}{\theta}d\varphi^{2}),
\end{displaymath}
where $r\geq 0,\,\theta\in [0,\pi],\,\varphi\in [0,2\pi].$
Asymptotic flatness is expressed by the boundary conditions
\begin{displaymath}
\lim_{r\rightarrow\infty}\lambda(r)=\lim_{r\rightarrow\infty}\mu(r)
=0,
\end{displaymath}
and a regular centre requires
\begin{displaymath}
\lambda(0)=0.
\end{displaymath}
Vlasov matter is described within the framework of kinetic theory. The fundamental object is the distribution function $f$ which is defined on phase-space, and models a collection of particles. The particles are assumed to interact only via the gravitational field created by the particles themselves and not via direct collisions between them. For an introduction to kinetic theory in general relativity and the Einstein-Vlasov system in particular we refer to \cite{An2} and \cite{Rl7}. 
The static Einstein-Vlasov system is given by the Einstein equations 
\begin{eqnarray}
&\displaystyle e^{-2\lambda}(2r\lambda_{r}-1)+1=8\pi r^2\rho,&\label{ee12}\\
&\displaystyle e^{-2\lambda}(2r\mu_{r}+1)-1=8\pi r^2
p,&\label{ee22}\\
&\displaystyle
\mu_{rr}+(\mu_{r}-\lambda_{r})(\mu_{r}+\frac{1}{r})= 4\pi
qe^{2\lambda},&\label{ee4}
\end{eqnarray}
together with the (static) Vlasov equation 
\begin{equation}
\frac{w}{\varepsilon}\partial_{r}f -(\mu_{r} \varepsilon-
\frac{L}{r^3\varepsilon})\partial_{w}f=0,\label{vlasov}
\end{equation}
where
\begin{equation*}
\varepsilon=\varepsilon(r,w,L)=\sqrt{1+w^{2}+L/r^{2}}.
\end{equation*}
The matter quantities are defined by
\begin{eqnarray*}
\rho(r)&=&\frac{\pi}{r^{2}}
\int_{-\infty}^{\infty}\int_{0}^{\infty}\varepsilon f(r,w,L)\;dLdw,\\
p(r)&=&\frac{\pi}{r^{2}}\int_{-\infty}^{\infty}\int_{0}^{\infty}
\frac{w^{2}}{\varepsilon}f(r,w,F)\;dLdw,\label{p}\\
q(r)&=&\frac{\pi}{r^{4}}\int_{-\infty}^{\infty}\int_{0}^{\infty}\frac{L}{\varepsilon}f(r,w,L)\;
dLdw. 
\end{eqnarray*}
The variables $w$ and $L$ can be thought of as the momentum in the radial direction 
and the square of the angular momentum respectively.
Let $$E=e^{\mu}\varepsilon,$$ the ansatz 
\begin{equation}
f(r,w,L)=\Phi(E,L), \label{ansats}
\end{equation}
then satisfies (\ref{vlasov}) and constitutes an efficient way to construct static 
solutions with finite ADM mass and finite extension, cf. \cite{RR4}, \cite{R1}. It should be pointed out that spherically symmetric static solutions which do not have this form globally exist, cf. \cite{Sc2}, which contrasts the Newtonian case where all spherically symmetric static solutions have the form (\ref{ansats}), cf. \cite{Ba}. As a matter of fact, the solutions we construct in Theorem 1 below are good candidates for solutions which are not globally given by (\ref{ansats}). 

Here the following form of $\Phi$ will be used 
\begin{equation}
\Phi(E,L)=(E_0-E)^k_{+}(L-L_0)_{+}^l,\label{pol}
\end{equation}
where $l\geq 1/2,\,k\geq 0,\,L_0>0,\; E_0>0,$ and $x_{+}:=\max\{x,0\}$.
In the Newtonian case with $l=L_0=0,$ this ansatz leads
to steady states with a polytropic equation of state. 
Note that when $L_{0}>0$ there will be no matter in the region
\begin{equation}
r<\sqrt{\frac{L_0}{(E_{0}e^{-\mu(0)})^2-1}},\label{r0}
\end{equation}
since there necessarily
$E>E_{0}$ and $f$ vanishes. The existence of solutions supported in $[R_0,R_1],\; R_0>0,$ with finite ADM mass has been given in \cite{R1}, and we shall call such configurations 
static shells of Vlasov matter. It will be assumed that $\Phi$ is always as above, which in particular means that $L_0>0,$ so that only shells are considered. The case $L_0=0$ is left to a future study.\\ 

Let the matter content within the sphere of area radius $r$ be defined by 
\[
m(r)=\int_0^r 4\pi\eta^2\rho d\eta. 
\]
Note that the total ADM mass $M=\lim_{r\to\infty}m(r).$ We also note that equation (\ref{ee12}) implies that 
\[
e^{-2\lambda}=1-\frac{2m(r)}{r},
\] 
so that $\mu$ alone can be regarded as the unknown metric function of the spherically symmetric Einstein-Vlasov system. 

Numerical evidence is presented in \cite{AR2} that the following hold true: \\
\newline
i) For any solution of the static Einstein-Vlasov system
\begin{equation}
\Gamma:=\sup_{r}\frac{2m(r)}{r}<\frac{8}{9}. \label{arb}
\end{equation}
ii) The inequality is sharp in the sense that there is a sequence of
steady states such that $\Gamma=2M/R=8/9$ in the limit. \\
\newline
These statements hold for both shells ($L_0>0$) and non-shells ($L_0=0$). More information on the latter case is given in \cite{AR2}. 
In the former case the sequence which realizes $\Gamma=8/9$ is obtained numerically in \cite{AR2} by constructing solutions where the inner boundary of the shells tend to zero. The outer boundary of these shells also tend to zero, and 
in \cite{AR2} numerical support is obtained for the following claim: \\ 
\newline
iii) There is a sequence of static shells supported in $[R_0^j,R_1^j],$ such that $R_0^j\to 0,$ and $R_1^j/R_0^j\to 1,$ as $j\to \infty.$ \\ 
\newline 
Our main results are described in the next section and concern issues (ii) and (iii) which will be proved. Of course, issue (i) is a very interesting open problem, and we believe that the results in this paper are important for proving also issue (i), in view of (ii). 

Let us end this section with a brief discussion on the possible role of the Buchdahl inequality for the time dependent problem with Vlasov matter. The cosmic censorship conjecture is fundamental in classical general relativity and to a large extent an open problem. In the case of gravitational collapse the only rigorous result is by Christodoulou who has obtained a proof in the case of the spherically symmetric Einstein-Scalar Field system \cite{Ch2}. One key result for this proof is contained in \cite{Ch1}, cf. also \cite{Ch3}, and states roughly that if there is a sufficient amount of matter within a bounded region then necessarily a trapped surface will form in the evolution. If a trapped surface forms, then Dafermos \cite{D} has shown under some restrictions on the matter model, that cosmic censorship holds. In particular, Dafermos and Rendall \cite{DR} have proved that spherically symmetric Vlasov matter satisfies these restrictions. Thus cosmic censorship holds for the spherically symmetric Einstein-Vlasov system if there is a trapped surface in spacetime. Now, assume that a Buchdahl inequality holds in general for this system, then in view of the result by Christodoulou mentioned above, it is natural to believe that if $2m/r$ exceeds the value given by such an inequality, then a trapped surface must form. 

The outline of the paper is as follows. 
In the next section our main results are presented in detail. Some preliminary results on static shells of Vlasov matter are contained in section 3, and in section 4 the proofs of the theorems are given. 
\section{Main results}
In view of (\ref{r0}) it is clear that the region where $f$ necessarily vanishes can be made arbitrary small if the values of $E_0$ and $\mu(0)$ are such that $E_0e^{-\mu(0)}$ is large. That this is always possible can be seen as follows. Set $E_0=1,$ and construct a solution by specifying an arbitrary non-positive value $\mu(0),$ in particular $e^{-\mu(0)}$ can be made as large as we wish. The metric function $\mu$ is then obtained by integrating from the centre using equation (\ref{ee22}), 
\begin{equation*}
\mu(r)=\mu(0)+\int_0^r (\frac{m}{r^2}+4\pi\eta p)e^{2\lambda}d\eta. 
\end{equation*}
This implies that the boundary condition at $\infty$ of $\mu$ will be violated in general. However, by letting $\tilde{E}_0=e^{\mu(\infty)},$ and $\tilde{\mu}(r):=\mu(r)-\mu(\infty),$ then $\tilde{\mu},$ and the distribution function $f$ associated with $\tilde{\mu}$ and $\tilde{E}_0,$ will solve the Einstein-Vlasov system and satisfy the boundary condition at infinity, and in view of (\ref{rho})-(\ref{q}), the matter terms will be identical to the original solution since $$\tilde{E}_0e^{\tilde{\mu}(r)}=e^{\mu(r)}.$$ Hence we will always take $E_0=1$ and obtain arbitrary small values of $R_0$ by taking $-\mu(0)$ sufficiently large. 

Let us define $$R_0:=\sqrt{\frac{L_0}{e^{-2\mu(0)}-1}}.$$ It will be clear from the proofs that in fact $f(r,\cdot,\cdot)>0,$ when $r$ is sufficiently close to but larger than $R_0.$ Hence, given any number $R_0>0$ we can construct a solution having inner radius of support equal to $R_0.$ 
The following result proves issue (iii). The constants $q, C_1, C_2$ and $C_3$ which appear in the formulation of the theorem are specified in equations (\ref{rhosim})-(\ref{zsim}). 
\begin{theorem}Consider a shell solution with a sufficiently small inner radius of support $R_0$. The distribution function $f$ then vanishes within the interval $$[R_0+B_0R_0^{(q+3)/(q+1)},(1-B_1R_0^{2/(q+1)})^{-1}(R_0+B_2 R_0^{(q+2)/(q+1)})],$$ 
where $B_0,B_1$ and $B_2$ are positive constants which depend on $C_1, C_2$ and $C_3.$ 
The solution can thus be joined with a Schwarzschild solution at the point where $f$ vanishes and a static shell is obtained with support within $[R_0,R_1],$ where $$R_1=\frac{R_0+B_2 R_0^{(q+2)/(q+1)}}{1-B_1R_0^{2/(q+1)}},$$ 
so that $R_1/R_0\to 1$ as $R_0\to 0.$  
\end{theorem}
This result is interesting in its own right since it gives a detailed description of the support of a class of shell solutions to the Einstein-Vlasov system. Moreover, the solutions constructed in Theorem 1 can be used to obtain a sequence of shells of Vlasov matter with the property that $2M/R=8/9$ in the limit. 
\begin{theorem}
Let $(f^{j},\mu^j)$ be a sequence of shell solutions with support in $[R_0^j,R_1^j],$ and such that $R_1^j/R_0^j\to 1$ and $R_0^j\to 0,$ as $j\to\infty,$ and let $M^j$ be the corresponding ADM mass of $(f^{j},\mu^j).$ 
Then 
\begin{equation}
\lim_{j\to\infty}\frac{2M^j}{R_1^j}=\frac{8}{9}.\label{limit2} 
\end{equation}
\end{theorem}
\section{Static shells of Vlasov matter}
When the distribution function $f$ has the form 
\begin{equation}
f(r,w,L)=\Phi(E,L),
\end{equation}
the matter quantities $\rho, p$ and $q$ become
functionals of $\mu$, and we have 
\begin{eqnarray}
\rho &=& \frac{2\pi}{r^2}\int_{\sqrt{1+\frac{L_0}{r^2}}}^{E_0e^{-\mu}}\int_{L_0}^{r^2(k^2-1)}\Phi(e^{\mu}k,L)\frac{k^2}{\sqrt{k^2-1-L/^2}}dLdk,\label{rho}\\
p &=& \frac{2\pi}{r^2}\int_{\sqrt{1+\frac{L_0}{r^2}}}^{E_0e^{-\mu}}\int_{L_0}^{r^2(k^2-1)}\Phi(e^{\mu}k,L)\sqrt{k^2-1-L/^2}dLdk,\label{p}\\
q &=& \frac{2\pi}{r^4}\int_{\sqrt{1+\frac{L_0}{r^2}}}^{E_0e^{-\mu}}\int_{L_0}^{r^2(k^2-1)}\Phi(e^{\mu}k,L)\frac{L}{\sqrt{k^2-1-L/^2}}dLdk.\label{q}
\end{eqnarray}
Here we have kept the parameter $E_0$ but recall that $E_0=1$ in what follows. 
If (\ref{pol}) is chosen for $\Phi$ these integrals can be computed explicitly 
in the cases when $k=0,1,2,...$ and $l=1/2,3/2,...$ as the following lemma shows. 
Let $$\gamma=-\mu-\frac{1}{2}\log{(1+\frac{L_0}{r^2})}.$$ 
\begin{lemma}Let $k=0,1,2,...$ and let $l=1/2,3/2,5/2,...$ then there are positive constants $\pi^{j}_{k,l},\, j=1,2,3$ such that when $\gamma\geq 0$
\begin{eqnarray}
\rho &=& \pi^{1}_{k,l}r^{2l}(1+\frac{L_0}{r^2})^{l+2}(e^{\gamma}-1)^{2l+k+1}P_{3-k}(e^{\gamma}),\\
p &=& \pi^{2}_{k,l}r^{2l}(1+\frac{L_0}{r^2})^{l+2}(e^{\gamma}-1)^{2l+k+2}P_{2-k}(e^{\gamma}),\\
z &=& \pi^{3}_{k,l}r^{2l}(1+\frac{L_0}{r^2})^{l+1}(e^{\gamma}-1)^{2l+k+1}P_{1-k}(e^{\gamma}).
\end{eqnarray}
If $\gamma<0$ then all matter components vanish. 
Here $P_{n}(e^{\gamma})$ is a polynomial of degree $n$ and $P_n>0,$ and $z:=\rho-p-q.$ 
\end{lemma}
\textit{Remark: }The restriction $l\geq 1/2$ is made so that the matter terms get the form above which is convenient. We believe however that the cases $0\leq l<1/2$ can be treated by similar arguments as presented below. 

\textit{Sketch of proof of Lemma 1: }To evaluate the $L-$integration in the expressions for $\rho$ and $z$ we substitute $x:=\sqrt{L-L_0}$ and use that for $n>0,$ 
\[
\int\frac{x^n}{\sqrt{ax^2+b}}=\frac{x^{n-1}\sqrt{ax^2+b}}{na}-\frac{(n-1)b}{na}\int\frac{x^{n-2}}{\sqrt{u}}dx.
\]
The same substitution is made for $p$ together with 
  \[
\int x^n\sqrt{ax^2+b}=\frac{x^{n-1}(ax^2+b)^{3/2}}{(n+2)a}-\frac{(n-1)b}{(n+2)a}\int x^{n-2}\sqrt{u}dx.
\]
The $k-$integration is straightforward and the claimed expressions follow by integration by parts and using that $$e^{\mu}=\frac{E_0e^{-\gamma}}{\sqrt{1+L_0/r^2}}.$$ 
\begin{flushright}
$\Box$
\end{flushright}
Next we show that if $R_0$ is small then also $\gamma$ is small. 
\begin{lemma}Given $k$ and $l$ there is a $c_{\gamma}>0$ and a $\delta_{\gamma}>0$ such that 
$$e^{\gamma(r)}-1\leq c_{\gamma}r^{2/(2l+k+2)}, \mbox{ for all }r\in [R_0,5R_0], \mbox{ when }
R_0\leq\delta_{\gamma}.$$
\end{lemma}

\textit{Proof of Lemma 2: }
Since $\gamma=0$ at $r=R_0$ we can consider an interval $I:=[R_0,y], y>R_0,$ such that $\gamma\leq 1$ on this interval. Hence on $I,$ $P_{2-k}(e^{\gamma})>C_P$ for some $C_P>0.$ We have 
\begin{eqnarray}
\gamma'(r)&=&-\mu'(r)+\frac{L_0}{r(r^2+L_0)}= -(\frac{m}{r^2}+4\pi rp)e^{2\lambda}+\frac{L_0}{r(r^2+L_0)}\nonumber\\
&\leq&-(\frac{m}{r^2}+4\pi r\frac{C(e^{\gamma}-1)^{2l+k+2}}{r^4})e^{2\lambda}+\frac{L_0}{r(r^2+L_0)}.\label{gammaprime}
\end{eqnarray} 
Here the constant $C$ depends on $C_P$ and $L_0.$ 
Since $\gamma(R_0)=0$ and $e^{2\lambda}\geq 1,$ this implies that $$(e^{\gamma}-1)^{2l+k+2}\leq\frac{r^2}{4\pi C},$$ since otherwise $\gamma'(r)\leq 0$ and $\gamma$ cannot increase while the right-hand side in this inequality is increasing in $r.$ It follows that $e^{\gamma(r)}\leq C r^{2/(2l+k+2)}+1,$ for $R_0\leq r\leq 5R_0,$ if $R_0$ is sufficiently small, since $\gamma$ is then less or equal to one on $[R_0,5R_0],$ which was an assumption of the argument. 
\begin{flushright}
$\Box$
\end{flushright}
\textit{Remark: }The assumption $\gamma\leq 1$ is not needed if $2-k\geq 0.$ The choice $[R_0,5R_0]$ was arbitrary and can be replaced by $[R_0,NR_0],\; N>0,$ by taking $R_0$ accordingly.

Since we will show that the interval of support $[R_0,R_1]$ is such that $R_1/R_0\to 1,$ as $R_0\to 0,$ the interval $[R_0,5R_0]$ in the lemma is no restriction and therefore we can always assume that $$e^{\gamma(r)}\leq Cr^{2/(2l+k+2)}+1.$$ 
Since $R_0$ is small this implies that $\gamma$ is small so that in particular $\gamma\leq e^{\gamma}-1\leq 2\gamma$ on $[R_0,5R_0].$ Thus, from Lemma 1 it follows that there are positive constants $C_U$ and $C_L$ such that 
\begin{equation}
C_L\frac{\gamma^{2l+k+1}}{r^4}\leq \rho \leq C_U\frac{\gamma^{2l+k+1}}{r^4},\label{ul}
\end{equation}
and analogously for $p$ and $z.$ For non-integer values on $k\geq 0$ and $l\geq 1/2,$ one can use the strategy described in the sketch of proof of Lemma 1 and obtain sufficient information to conclude that the claim in Lemma 2, i.e. that $\gamma$ is small whenever $r$ is small, holds also for non-integer values of $k$ and $l.$ It is then straightforward to make upper and lower estimates on the matter terms with respect to the corresponding matter terms for integer values on $k$ and $l.$ Therefore, since upper and lower estimates as in (\ref{ul}) are sufficient for all the arguments below we will for simplicity, and without loss of generality, assume that for some positive constants $C_1, C_2,$ and $C_3,$ 
\begin{eqnarray}
\rho &=& C_1\frac{\gamma^{q}}{r^4},\label{rhosim}\\
p &=& C_2\frac{\gamma^{q+1}}{r^4},\label{psim}\\
z &=& C_3\frac{\gamma^{q}}{r^2},\label{zsim}
\end{eqnarray}
where $q=2l+k+1\geq 2.$ 
\section{Proofs of Theorems 1 and 2}
\textit{Proof of Theorem 1: }
We will always take $R_0\leq 1$ sufficiently small so that the statement in Lemma 2 holds and so that  
\begin{equation}
\frac{L_0}{r^2+L_0}\geq 5/6, \mbox{ for } r\in [R_0,5R_0].\label{L0} 
\end{equation}
It will also be tacitly assumed that all intervals we consider below are subsets of $[R_0,5R_0].$ It will be clear from the arguments that this can always be achieved by taking $R_0$ sufficiently small. The positive constant $C$ can change value from line to line. 
The proof of Theorem 1 will follow from a few lemmas and a proposition. 
\begin{lemma}
Let $C_m=\max{\{1,C_1,4\pi C_2\}}$ and let $$\delta\leq \frac{R_0^{(q+3)/(q+1)}}{C_m8^{1/q+1}},$$ then $$\gamma'(r)\geq\frac{1}{2r},$$ for 
$r\in [R_0,R_0+\delta].$ 
\end{lemma}
\textit{Proof of Lemma 3: }We have 
$$\gamma'(r)=-\mu'(r)+\frac{L_0}{r(r^2+L_0)}.$$ Let $\sigma\in [0,\delta],$ since $\gamma(R_0)=0,$ it follows that 
\begin{equation}
\gamma(R_0+\sigma)=\gamma(R_0+\sigma)-\gamma(R_0)\leq\sigma\gamma'(\xi)\leq\frac{\delta}{R_0},\label{gammaleq}
\end{equation}
where $\xi\in [R_0,R_0+\sigma].$ 
Hence, by (\ref{rhosim}) we get 
$$\rho\leq\frac{C_1\delta^q}{r^4R_0^q}, \mbox{ for }r\in [R_0,R_0+\delta].$$ 
Using that $\rho=0$ when $r<R_0,$ we obtain for $\sigma\in [0,\delta],$
\[
m(R_0+\sigma)\leq\int_{R_0}^{R_0+\sigma}\frac{C_1\delta^q}{\eta^4R_0^q}\eta^2 d\eta=\frac{C_1\delta^q \sigma}{R_0^{q+1}(R_0+\sigma)}\leq \frac{C_1\delta^{q+1}}{R_0^{q+1}(R_0+\sigma)}.
\]
Hence, 
\[
\frac{m(R_0+\sigma)}{R_0+\sigma}\leq\frac{C_1\delta^{q+1}}{R_0^{q+1}(R_0+\sigma)^2}\leq \frac{C_1\delta^{q+1}}{R_0^{q+3}}.
\]
From (\ref{psim}) we also have 
$$p(r)\leq\frac{C_2\delta^{q+1}}{r^4R_0^{q+1}}, \mbox{ for }r\in [R_0,R_0+\delta].$$ 
Note that by taking $\delta^{q+1}\leq R_0^{q+3}/8C_m,$ we have $m(R_0+\sigma)/(R_0+\sigma)\leq 1/8$ so that $e^{2\lambda(R_o+\sigma)}\leq 4/3,$ and 
$$4\pi r^2 p\leq \frac{R_0^{q+3}}{8r^2R_0^{q+1}}\leq \frac{1}{8}.$$ Thus for $r\in [R_0,R_0+\delta],$ 
$$r\mu'(r)=\frac{m(r)}{r}e^{2\lambda(r)}+4\pi r^2 p(r)e^{2\lambda(r)}\leq 1/6+1/6=1/3.$$
In view of (\ref{L0}) we thus have $$\gamma'(r)\geq -\frac{1}{3r}+ \frac{L_0}{r(r^2+L_0)}\geq 1/2r,$$ for $r\in [R_0,R_0+\delta],$ and the lemma follows. 
\begin{flushright}
$\Box$
\end{flushright}
The lemma implies that for $\sigma\in [0,\delta],$ 
\begin{equation}
\gamma(R_0+\sigma)\geq\gamma(R_0)+\sigma\inf_{\sigma\in [0,\delta] }\gamma'(R_0+\sigma)\geq\frac{\sigma}{2(R_0+\sigma)},\label{gammalower}
\end{equation}
where we again used that $\gamma(R_0)=0.$ Let $$\sigma_{*}:=C_{0}R_0^{1+\frac{2q+1}{q(q+1)}},$$ where $C_0$ satisfies the conditions $C_0\leq 1/4$ and $C_0\leq 1/(C_m8^{1/(q+1)}).$  
Now, since $$1+\frac{2q+1}{q(q+1)}=\frac{q+3}{q+1}+\frac{1}{q(q+1)}\geq\frac{q+3}{q+1},$$ we have in view of the second assumption on $C_0$ that 
$$\sigma_{*}<\delta.$$ 
Define $\gamma_{*}$ by 
$$\gamma_{*}=\frac{\sigma_{*}}{2(R_0+\sigma_{*})}.$$ 
It is clear that $\gamma(R_0+\sigma_{*})\geq\gamma_{*}$ in view of (\ref{gammalower}). We will show that $\gamma$ must reach the $\gamma_{*}-$level again at $r_2$ (i.e. a second time) close to $R_0+\sigma_{*}.$ We point out that the choice of the exponent in the definition of $\sigma_*$ is crucial, and our arguments provide almost no room to choose it differently. We have 
\begin{lemma}Let $\kappa=3^22^{2q}/(C_1C_0^q)$ and let $\Gamma=\max{\{C_0,\kappa\}},$ and consider a solution with $R_0$ such that $R_0^{1/(q+1)}\Gamma\leq 1.$ Then there is a point $r_2$ such that $\gamma(r)\downarrow\gamma_{*}$ as $r\uparrow r_2,$ and $r_2\leq R_0+\sigma_{*}+\kappa R_0^{(q+2)/(q+1)}.$ 
\end{lemma}
\textit{Proof of Lemma 4. }Since $\gamma(R_0+\sigma_{*})\geq\gamma_{*},$ and since Lemma 3 gives that $\gamma'(R_0+\sigma_{*})>0,$ the point $r_2$ must be strictly greater than $R_0+\sigma_{*}.$ Let $[R_0+\sigma_{*},R_0+\sigma_{*}+\Delta],$ for some $\Delta>0,$ be such that $\gamma\geq\gamma_{*}$ on this interval. We will show that 
\[
\Delta\leq\kappa R_0^{(q+2)/(q+1)}. 
\] 
We have from (\ref{rhosim}) 
\begin{eqnarray}
\displaystyle m(r)&\geq& C_1\int_{R_0+\sigma_{*}}^{R_0+\sigma_{*}+\Delta}\frac{\sigma_{*}^q}{r^4 2^{q}(R_0+\sigma_{*})^q}r^2dr\nonumber \\
\displaystyle &=&C_1\frac{\sigma_{*}^q}{2^{q}(R_0+\sigma_{*})^{q+1}}\frac{\Delta}{(R_0+\sigma_{*}+\Delta)}.\label{msigma}
\end{eqnarray}
Hence, 
$$\frac{m(R_0+\sigma_{*}+\Delta)}{R_0+\sigma_{*}+\Delta}\geq C_1\frac{\sigma_{*}^q}{2^q(R_0+\sigma_{*})^{q+1}}\frac{\Delta}{(R_0+\sigma_{*}+\Delta)^2}.$$ 
The assumption in the lemma guarantees that $\sigma_{*}\leq R_0$ since 
\[
\frac{2q+1}{q(q+1)}\geq\frac{1}{q+1}.
\] 
Substituting for $\sigma_{*}$ then gives 
\begin{eqnarray*}
\frac{m(R_0+\sigma_{*}+\Delta)}{R_0+\sigma_{*}+\Delta}&\geq& C_1C_0^q\frac{R_0^{q\left(1+\frac{2q+1}{q(q+1)}\right)}}{2^{2q+1}R_0^{q+1}}\frac{\Delta}{(R_0+\sigma_{*}+\Delta)^2}\\
&=& C_1C_0^q\frac{R_0^{q/(q+1)}\Delta}{2^{2q+1}(R_0+\sigma_*+\Delta)^2}. 
\end{eqnarray*}
Let $\Delta=\kappa R_0^{(q+2)/(q+1)},$ and note that the assumptions 
of the lemma imply that $\Delta\leq R_0.$ It follows that 
\begin{equation*}
\frac{m(R_0+\sigma_{*}+\Delta)}{R_0+\sigma_{*}+\Delta}\geq C_1C_0^q\kappa\frac{R_0^{2}}{2^{2q+1}(3R_0)^2}=\frac{C_1C_0^q\kappa}{3^22^{2q+1}}.
\end{equation*} 
The definition of $\kappa$ implies that 
$$\frac{m(R_0+\sigma_{*}+\Delta)}{R_0+\sigma_{*}}\geq \frac{1}{2}.$$ This is impossible since it is proved in \cite{BR} that all static solutions have $2m/r<1,$ and therefore $r_2$ must be strictly less than $R_0+\sigma_{*}+\kappa R_0^{(q+2)/(q+1)}.$ 
This completes the proof of the lemma. 
\begin{flushright}
$\Box$
\end{flushright}
We will next show that if $R_0$ is sufficiently small then $2m/r$ will attain values arbitrary close to $8/9.$ 
\begin{proposition}
Let $r_2$ be as in Lemma 4 for a sufficiently small $R_0.$ Then the corresponding solution satisfies $m(r_2)/r_2\geq 2/5,$ and if $R_0\to 0,$ then $2m(r_2)/r_2\to 8/9.$ 
\end{proposition} 
\textit{Proof of Proposition 1: }
We now consider the fundamental equation ($10$) in \cite{An1} which reads 
\begin{equation}
(\frac{m}{r^2}+4\pi rp)e^{\mu+\lambda}=\frac{1}{r^2}\int_0^r 4\pi\eta^2 e^{\mu+\lambda}(\rho+p+q)d\eta.\label{fundamentaleq}
\end{equation}
This equation a consequence of the generalized (for non-isotropic pressure) Oppenheimer-Tolman-Volkov equation. For $r=r_2$ we then have 
\begin{eqnarray}
\displaystyle m(r_2)e^{(\mu+\lambda)(r_2)}&=&\int_{R_0}^{r_2} 4\pi\eta^2 e^{\mu+\lambda}(\rho+p+q)\,d\eta-4\pi r_2^3pe^{(\mu+\lambda)(r_2)}\nonumber\\ 
\displaystyle &=& \int_{R_0}^{r_2}4\pi\eta^2 e^{\mu+\lambda}(2\rho-z)\,d\eta-4\pi r_2^3pe^{(\mu+\lambda)(r_2)}.\label{Mr2inequality} 
\end{eqnarray}
Here we used that $p+q=\rho-z.$ From (\ref{zsim}) and (\ref{rhosim}) we get 
\begin{eqnarray}
\int_{R_0}^{r_2}4\pi\eta^2 e^{\mu+\lambda}z\,d\eta&\leq& \frac{C_3}{C_1}\int_{R_0+\sigma_{*}}^{r_2}4\pi\eta^4\gamma\rho e^{\mu+\lambda}\,d\eta\nonumber\\
&\leq&\frac{C_3r_2^3}{C_1}\int_{R_0+\sigma_{*}}^{r_2}4\pi\eta\rho e^{\mu+\lambda}\,d\eta\nonumber\\
&\leq&Cr_2^3e^{\mu(r_2)}\int_{R_0+\sigma_{*}}^{r_2}4\pi\eta\rho e^{\lambda}\,d\eta.\label{zestimate}
\end{eqnarray}
Here we used that $e^{\mu}$ is increasing. 
Next we observe that the result of Theorem 1 in \cite{An1} can be applied to the interval $[R_0,r_2]$ when $R_0$ is small enough. Indeed, replace $M$ by $m(r_2),$ use the fact that $e^{\mu(r_2)}\leq e^{-\lambda(r_2)}=\sqrt{1-2m(r_2)/r_2},$ and use the fact that $p(r_2)\geq 0$ so that the second term on the left hand side of equation (\ref{fundamentaleq}) can be dropped. Therefore, since $r_2/R_0\to 1,$ Theorem 1 in \cite{An1} implies that for a sufficiently small $R_0$ there exists a positive number $k,$ less than one, such that 
\[
\sup_{r\in [R_0,r_2]}\frac{2m(r)}{r}\leq 1-k, 
\]
so that $\lambda\leq -\log{\sqrt{k}}=:C_{\lambda}$ on the interval $[R_0,r_2].$ (Note also that $C_{\lambda}$ can be made arbitrary close to $\log{3}$ by Theorem 2 in \cite{An1} by taking $R_0$ sufficiently small.) 
Now we write 
\begin{eqnarray}
&\displaystyle\int_{R_0}^{r_2}4\pi\eta\rho e^{\lambda}\,d\eta=\int_{R_0}^{r_2}(-\frac{d}{dr}e^{-\lambda})d\eta+\int_{R_0}^{r_2}\frac{me^{\lambda}}{\eta^2}d\eta&\nonumber\\
&\displaystyle\leq 1-\sqrt{1-\frac{2m(r_2)}{r_2}}+e^{C_{\lambda}}m(r_2)\frac{r_2-R_0}{r_2R_0}&\nonumber\\
&\displaystyle\leq 1-\sqrt{1-\frac{2m(r_2)}{r_2}}+\frac{Cm(r_2)R_0^{(q+2)/(q+1)}}{R_0^{2}}&\\
&\displaystyle =\frac{2m(r_2)}{r_2(1+\sqrt{1-2m(r_2)/r_2})}+\frac{Cm(r_2)}{R_0^{q/(q+1)}}&\nonumber\\
&\displaystyle\leq\frac{2m(r_2)}{R_0}+\frac{Cm(r_2)}{R_0^{q/(q+1)}}\leq\frac{Cm(r_2)}{R_0}.&\label{fundcomp}
\end{eqnarray}
Here we used that $r_2\leq R_0+\sigma_*+\kappa R_0^{(q+2)/(q+1)},$ and that $\sigma_*\leq CR_0^{(q+2)/(q+1)},$ for some $C>0$ when $R_0$ is small. 
From Lemma 4 we have that $\gamma$ approaches $\gamma_{*}$ from above and therefore $\gamma'(r_2)\leq 0,$ which implies that $\mu'(r_2)\geq 5/(6r_2),$ in view of (\ref{gammaprime}) and (\ref{L0}). Now, since $$\mu'=(\frac{m}{r^2}+4\pi rp)e^{2\lambda},$$ and since (\ref{psim}) gives $$r_2p(r_2)=C_2r_2\frac{\gamma_{*}^{q+1}}{r_2^4}\leq 
\frac{CR_0^{(2q+1)/q}}{r_2^{3}}\leq\frac{C}{R_0^{(q-1)/q}},$$
we necessarily have $m(r_2)/r_2\geq 1/4$ if $R_0$ is small enough. Indeed, if $m(r_2)/r_2\leq 1/4,$ then $e^{2\lambda}\leq 2,$ and we get $$\mu'(r_2)=(\frac{m}{r^2}+4\pi rp)e^{2\lambda}\leq \frac{1}{2r_2}+\frac{C}{R_0^{(q-1)/q}},$$ and this is smaller than $5/(6r_2)$ when $r_2,$ or equivalently $R_0,$ is small. 
Hence, $r_2\leq 4m(r_2),$ and from (\ref{Mr2inequality}), (\ref{zestimate}) and (\ref{fundcomp}) we obtain (using $R_0\leq r_2\leq 3R_0$) 
\begin{eqnarray}
&\displaystyle m(r_2)e^{(\mu+\lambda)(r_2)}\geq 2\int_{R_0}^{r_2}4\pi\eta^2 e^{\mu+\lambda}\rho\,d\eta -Ce^{\mu(r_2)}r_2^{2}m(r_2)&\nonumber\\
&\displaystyle -Cr_2^2m(r_2)pe^{(\mu+\lambda)(r_2)}\geq 2e^{\mu(R_0)}R_0
\int_{R_0}^{r_2}4\pi\eta e^{\lambda}\rho\,d\eta&\nonumber\\
&\displaystyle -Ce^{\mu(r_2)}R_0^{2}m(r_2)-CR_0^2m(r_2)pe^{(\mu+\lambda)(r_2)}.&\label{Mr2inequalityz} 
\end{eqnarray}
Here we again used that $e^{\mu}$ is increasing. 
For the integral term of the right hand side we use the computation in (\ref{fundcomp}), but now we estimate from below (since the sign is the opposite) and thus we do not need to estimate the integral 
$$\int_{R_0}^{r_2}\frac{me^{\lambda}}{\eta^2}d\eta,$$ which we drop and we get 
\begin{equation}
\int_{R_0}^{r_2}4\pi\eta e^{\lambda}\rho\,d\eta\geq \frac{2m(r_2)}{r_2(1+\sqrt{1-2m(r_2)/r_2})}. 
\end{equation}
Thus we obtain 
\begin{eqnarray}
m(r_2)e^{(\mu+\lambda)(r_2)}&\geq& e^{\mu(R_0)}\left(\frac{R_0}{r_2}\right)\frac{4m(r_2)}{1+\sqrt{1-2m(r_2)/r_2}}\nonumber\\
&-&Ce^{\mu(r_2)}R_0^{2}m(r_2)-CR_0^2m(r_2)pe^{(\mu+\lambda)(r_2)}.
\end{eqnarray}
Now we use again that $$p(r_2)=C_2\frac{\gamma_{*}^{q+1}}{r_2^4}\leq \frac{CR_0^{(2q+1)/q}}{r_2^{4}}\leq\frac{C}{R_0^{(2q-1)/q}},$$ together with the fact that $\lambda\geq 1,$ and obtain 
\begin{equation}
1\geq e^{\mu(R_0)-\mu(r_2)}\left(\frac{R_0}{r_2}\right)\frac{4e^{-\lambda(r_2)}}{1+\sqrt{1-2m(r_2)/r_2}}-CR_0^{2}-CR_0^{1/q}.\label{eqnbuchdahl}
\end{equation}
Let us now consider $\mu(R_0)-\mu(r_2).$ Since 
\[
\mu(r)=\mu(0)+\int_0^r (\frac{m}{\eta^2}+4\pi\eta p)e^{2\lambda}\,d\eta, 
\]
we have 
\[
\mu(R_0)-\mu(r_2)=-\int_{R_0}^{r_2}(\frac{m}{\eta^2}+4\pi\eta p)e^{2\lambda}\,d\eta. 
\]
We will show that the integral goes to zero as $R_0\to 0.$ From Lemma 2 we have that $\gamma\leq Cr^{2/(q+1)}$ (where we use the fact that $r$ is small so that at least $e^{\gamma}-1\leq 2\gamma$) and it follows by (\ref{psim}) that 
\[
p\leq \frac{C}{r^2}. 
\]
Since $\lambda\leq C_{\lambda}$ on $[R_0,r_2],$ and $m/r\leq 1/2$ always, we get 
\[
(\frac{m}{r^2}+4\pi rp)e^{2\lambda}\leq\frac{P}{r},
\]
where $P$ is a constant depending on $C_{\lambda}$ and $C_2.$ 
Hence, 
\begin{equation}
\mu(R_0)-\mu(r_2)\geq -P\log{(r_2/R_0)}. 
\end{equation}
This estimate implies that (\ref{eqnbuchdahl}) can be written
\begin{equation}
1\geq \left(\frac{R_0}{r_2}\right)^{P+1}\frac{4\sqrt{1-2m(r_2)/r_2}}{1+\sqrt{1-2m(r_2)/r_2}}-CR_0^{2}-CR_0^{1/q}.\label{eqnbuchdahl2}
\end{equation}
Since $R_0/r_2\uparrow 1,$ as $R_0\to 0,$ we can write this inequality as 
\[
1\geq \left(1-\Gamma(R_0)\right)\frac{4\sqrt{1-2m(r_2)/r_2}}{1+\sqrt{1-2m(r_2)/r_2}}-C\Gamma(R_0),
\]
where $\Gamma(R_0)\downarrow 0$ as $R_0\to 0.$ 
This yields 
\[
1+\sqrt{1-2m(r_2)/r_2}\geq \left(1-\Gamma(R_0)\right)4\sqrt{1-2m(r_2)/r_2}-C\Gamma(R_0),
\]
so that
\[
1\geq 3\sqrt{1-2m(r_2)/r_2}-C\Gamma(R_0).
\]
Squaring both sides and solving for $2m(r_2)/r_2$ gives 
\begin{equation}
\frac{2m(r_2)}{r_2}\geq \frac{8}{9}-C\Gamma(R_0).\label{geq} 
\end{equation}
It is now clear that $m(r_2)/r_2\geq 2/5,$ 
when $R_0$ is sufficiently small, and that $2m(r_2)/r_2\to 8/9$ as $R_0\to 0,$ which completes the proof of the proposition. 
\begin{flushright}
$\Box$
\end{flushright}
We can now finish the proof of Theorem 1. First of all note that $f$ cannot vanish for 
$$r\leq R_0+\frac{R_0^{(q+3)/(q+1)}}{C_m8^{1/(q+1)}},$$ 
by Lemma 3. Thus the claim that $f$ will not vanish before $r=R_0+B_0R_0^{(q+3)/(q+1)}$ follows with $B_0:=1/C_m8^{1/(q+1)}.$ 
The main issue is of course to prove that $f$ vanishes before 
\[
\frac{R_0+B_2R_0^{(q+2)/(q+1)}}{1-B_1R_0^{(q+2)/(q+1)}}, 
\] 
where $B_1$ and $B_2$ are positive constants. 
Inspired by an idea of T. Makino introduced in \cite{Ma}, we show that $\gamma$ necessarily must vanish close to the point $r_2$ if $R_0$ is sufficiently small. 
Let $$x:=\frac{m(r)}{r\gamma(r)}.$$
Using that $m'(r)=4\pi r^2\rho,$ it follows that 
\[
rx'=\frac{4\pi r^2\rho}{\gamma}-x+\frac{x^2}{1-2\gamma x}-\frac{xL_0}{\gamma(r^2+L_0)}.
\]
In our case $r>R_0$ and $\gamma>0$ and we will show that $\gamma(r)=0$ for some $r<(1+\Gamma(R_0))r_2,$ where $\Gamma$ has the property as in the proof of Proposition 1. Since $\gamma>0$ and $\rho\geq 0$ the first term can be dropped and we have 
\begin{equation}
rx'\geq -x+\frac{x^2}{1-2\gamma x}-\frac{xL_0}{\gamma(r^2+L_0)}=\frac{x^2}{3(1-2\gamma x)}-x+\frac{2x^2}{3(1-2\gamma x)}-\frac{xL_0}{\gamma(r^2+L_0)}. 
\end{equation}
Take $R_0$ sufficiently small so that $m(r_2)/r_2\geq 2/5$ by Proposition 1. Let 
$r\in [r_2,16r_2/15],$ then since $m$ is increasing in $r$ we get 
\[
\frac{m(r)}{r}\geq\frac{m(r_2)}{r}=\frac{r_2}{r}\frac{m(r_2)}{r_2}\geq\frac{15}{16}\cdot\frac{2}{5}=\frac{3}{8}.\]
Now by the definition of $x$ it follows that 
$$\frac{x}{1-2\gamma x}=\frac{m}{\gamma r}e^{2\lambda}=\frac{m}{\gamma r(1-2m/r)}\geq\frac{3}{2\gamma}, \mbox{ when } \frac{m}{r}\geq\frac{3}{8}.$$
Thus on $[r_2,16r_2/15],$
$$\frac{2x^2}{3(1-2\gamma x)}-\frac{xL_0}{\gamma(r^2+L_0)}\geq 0,$$
so that on this interval 
\begin{equation}
rx'\geq\frac{x^2}{3(1-2\gamma x)}-x\geq\frac{4}{3}x^2-x,\label{makino} 
\end{equation}
where we used that 
$$\frac{1}{1-2\gamma x}=\frac{1}{1-2m/r}\geq 4 \mbox{ when } \frac{m}{r}\geq\frac{3}{8}.$$
Lemma 2 gives an upper bound of $\gamma$ which implies that 
\begin{equation}
x(r_2)=\frac{m(r_2)}{r_2\gamma(r_2)}\geq\frac{2}{5}\cdot\frac{C}{r_2^{2/(q+1)}}.\label{xr2}
\end{equation}
Thus $x(r_2)\to\infty$ as $R_0\to 0,$ and we take $R_0$ sufficiently small so that 
$$\frac{x(r_2)}{x(r_2)-3/4}\leq\frac{16}{15}.$$ In particular $x(r_2)\geq 3/4.$ 
Solving (\ref{makino}) yields 
$$x(r)\geq\left(1-\frac{r(4x(r_2)/3-1)}{4r_2x(r_2)/3}\right)^{-1}, \mbox{ on } r\in [r_2,16r_2/15),$$
and we get that $x(r)\to\infty$ as $r\to R_1,$ where 
\begin{equation}
R_1\leq r_2\frac{x(r_2)}{x(r_2)-3/4}\leq\frac{16r_2}{15}.\label{R1}
\end{equation}
Now in view of (\ref{xr2}), 
$$\frac{x(r_2)}{x(r_2)-3/4}\leq\frac{1}{1-B_1R_0^{2/(q+1)}}\to 1, \mbox{ as }R_0\to 0,$$ for some positive constant $B_1.$ Since $\sigma_*\leq\kappa R_0^{(q+2)/(q+1)},$ if $R_0$ is sufficiently small, it is clear that there is a positive constant $B_2$ such that $r_2\leq R_0+B_2R_0^{(q+2)/(q+1)},$ and the proof of Theorem 1 is complete. 
\begin{flushright}
$\Box$
\end{flushright}
\textit{Proof of Theorem 2: }
From Theorem 2 in \cite{An1} we get with $\Omega=1$ that 
$$\limsup_{R_0\to 0}\frac{2M}{R_1}\leq\frac{8}{9}.$$
The arguments in Propostion 1 leading to (\ref{geq}) can be applied when $r=R_1$ instead of $r=r_2.$ Thus we also have 
$$\liminf_{R_0\to 0}\frac{2M}{R_1}\geq\frac{8}{9},$$ 
and the claim of the theorem follows. 
\begin{flushright}
$\Box$
\end{flushright}
\begin{center}
\textbf{Acknowledgement}
\end{center}
I want to thank Gerhard Rein for discussions and for commenting the manuscript, and Alan Rendall for drawing my attention to the Buchdahl inequality in connection to a result by Christodoulou \cite{Ch1} on the formation of trapped surfaces.

\end{document}